# A Marginal Reliability Impact Based Accreditation Framework for Capacity Markets

Feng Zhao, *Senior Member*, *IEEE*, Tongxin Zheng, *Fellow*, *IEEE*, Dane Schiro, *Member, IEEE*, Xiaochu Wang, *Member, IEEE*

*Abstract*— This paper presents a Marginal Reliability Impact (MRI) based resource accreditation framework for capacity market design. Under this framework, a resource is accredited based on its marginal impact on system reliability, thus aligning the resource's accreditation value with its reliability contribution. A key feature of the MRI-based accreditation is that the accredited capacities supplied by different resources to the capacity market are substitutable in reliability contribution, a desired feature of homogeneous products. Moreover, with MRI-based capacity demand, substitutability between supply and demand for capacity is also achieved. As a result, a capacity market with the MRI-based capacity product can better characterize the underlying resource adequacy problem and lead to more efficient market outcomes.

## I. INTRODUCTION

Adequacy is a key aspect of power system reliability. According to the North American Electric Reliability Corporation (NERC), adequacy is "the ability of the electricity system to supply the aggregate electrical demand and energy requirements of the end-use customers at all times, taking into account scheduled and reasonably expected unscheduled outages of system elements" [1]. Currently, several Regional Transmission Organizations (RTOs) in the US, e.g., ISO New England, Mid-continent ISO (MISO), New York ISO (NYISO) and PJM, use capacity markets to address their regions' adequacy needs. Resource accreditation is an important part of the capacity market design as it dictates the capacity product and determines the quantity that a resource can offer into the market. Installed Capacity (ICAP) and its outage rate discounted Unforced Capacity (UCAP) were often used as the accredited capacity in regional capacity markets [2].

While ICAP and UCAP have been adopted in conventional thermal dominated systems, extending these concepts to intermittent and energy storage resources with the grid transitioning to clean energy raises the issue of substitutability between different resource types. Therefore, in recent years many US regions have pursued capacity accreditation reforms [3]-[10]. ISO-NE is currently undergoing the transition from Qualified Capacity (QC) accreditation, an ICAP alike concept, to a Marginal Reliability Impact (MRI) based accreditation framework [3]; MISO has changed from UCAP to a Direct Loss of Load (DLOL) method, which resembles a marginal method, for their capacity accreditation [4]; NYISO has implemented a marginal reliability improvement based accreditation to replace UCAP for non-thermal resources [5]; and PJM has shifted from the UCAP accreditation to an average Effective Load Carrying Capability (ELCC) accreditation and more recently, a marginal ELCC method [6]. Compared to ICAP and UCAP, these new accreditations are intended to more accurately reflect a resource's contribution to system *adequacy* or *reliability*[1]. A shared feature of these methods is that a resource's accreditation value is affected not only by its own characteristics but by other resources as well, a result of system reliability *not* being an additively separable function of individual resources. Such interdependence allows these new accreditation methods to reflect the diversity impact of the resource mix, e.g., adding resources of the same characteristics tends to reduce their *marginal* reliability benefits.

This paper introduces the MRI-based accreditation framework that has served as a foundation for ISO-NE's ongoing capacity market reform [7]-[8]. Starting with the desired substitutability of a homogeneous capacity product, we first derive the general idea of MRI-based accreditation. By choosing Expected Unserved Energy (EUE) as the adequacy metric and perfect capacity as the reference, we then define MRI Capacity (MRIC) as a resource's accredited capacity. In conjunction with capacity market design, various aspects of the MRI-based accreditation framework, including *construction of MRIC demand curves, calculation of MRI for different resource types, interpretations of MRI and MRIC, its properties,* and *its alignment with the adequacy problem*, are then discussed. Furthermore, we compare the MRI-based accreditation to ICAP, UCAP, average and marginal ELCC approaches [11]-[12]. The main contributions of this paper are: (i) introducing a rigorous MRI-based accreditation framework for capacity market design; (ii) analyzing the interpretations and properties of the MRI-based accreditation; and (iii) providing a comprehensive comparison of different accreditation methods.

The rest of the paper is organized as follows. Section II provides details of the MRI-based accreditation framework. Section III compares MRI-based accreditation with other accreditation methods. Section IV summarizes numerical

---

testing results. Section V concludes the paper.

## II. THE MRI-BASED ACCREDITATION FRAMEWORK FOR CAPACITY MARKETS

The key idea of MRI-based accreditation is to align a resource's accredited capacity, i.e., the quantity the resource can offer into a capacity market, with its contribution to system adequacy. The MRI concept plays a central role in this alignment. The following Section II.A starts with the desired substitutability feature for a homogeneous capacity product to deduce the MRI and MRI Capacity (MRIC) concepts. Section II.B then derives the MRIC demand curves from our previously developed capacity demand curves [13]-[15]. Section II.C presents the calculation of MRI for different resource types. Section II.D explores the interpretations of MRI and MRIC. Section II.E examines the properties of MRI-based accreditation. Lastly, Section II.F provides insights into the potential market efficiency gain by aligning resources' accreditation values with their contributions to adequacy.

*II.A MRI-based Capacity Accreditation*

In capacity market design, a homogeneous capacity product is highly desirable. Homogeneity implies *substitutability*, i.e., each unit of accredited capacity, regardless of its source, provides the same benefit to system adequacy or reliability, or mathematically,

$$\frac{\partial M}{\partial \hat{C}_i} = -K, \quad \forall \text{ resource } i. \tag{1}$$

where $M$ is an adequacy risk metric[2], $\hat{C}_i$ is resource $i$'s *accredited capacity* or *market capacity*, and $K$ is a positive constant. The negative sign indicates reduced adequacy risk with the increase of capacity. Note that $M$ is a multi-variate function of all resources' capacities since system adequacy is affected by all resources in the system. Eq. (1) indicates that different resources' accredited capacities have the same marginal reliability contribution. Next, we derive the mathematical expression of $\hat{C}_i$ that satisfies (1).

Let $C_i$ be a parameter representing resource $i$'s own size, e.g., nameplate capacity, operating capability, etc. Hereafter we use *native capacity* or *physical capacity* for $C_i$ as it reflects resource $i$'s own physical characteristics. The native capacities of different resources may not be substitutable, e.g., one MW nameplate capacity of a nuclear resource could have a very different impact than that of a solar resource on system reliability. To have an accreditation scheme that satisfies the substitutability property (1), resource $i$'s native capacity $C_i$ is adjusted by a factor $\alpha_i \geq 0$ to define its accredited capacity $\hat{C}_i$, i.e.,

$$\hat{C}_i \equiv C_i \cdot \alpha_i, \quad \forall \text{ resource } i. \tag{2}$$

Substituting (2) into (1), we have $\frac{\partial M}{\partial \hat{C}_i} = \frac{\partial M}{\partial C_i} \cdot \frac{1}{\alpha_i} = -K$, or

$$\alpha_i = -\frac{\partial M}{\partial C_i} \cdot \frac{1}{K}, \forall \text{ resource } i. \tag{3}$$

Substituting (3) into (2), we obtain the following form of accredited capacity $\hat{C}_i$ that is substitutable in terms of marginal reliability contribution:

$$\hat{C}_i = C_i \cdot \left(-\frac{\partial M}{\partial C_i}\right) \cdot \frac{1}{K}, \quad \forall \text{ resource } i. \tag{4}$$

Eq. (4) shows that resource $i$'s accredited capacity $\hat{C}_i$, in MW, is the resource's native capacity $C_i$, in MW, adjusted by its marginal reliability impact $\left(-\frac{\partial M}{\partial C_i}\right)$, and normalized by a constant $K$ that has the same unit as the marginal reliability impact (in order to yield the accredited capacity $\hat{C}_i$ in MW). Note that the higher a resource's marginal reliability impact, the higher its accreditation value. This way, a resource's accredited capacity $\hat{C}_i$ accounts for its Marginal Reliability Impact (MRI) and is thus termed MRI Capacity (MRIC). With substitutability, MRIC can be treated as a homogeneous capacity product and priced uniformly in capacity market, allowing more market transparency and efficiency.

Below we discuss the choices of $M$ and $K$ in (4). In [16], NERC defines several system adequacy metrics such as Loss of Load Expectation (LOLE), Loss of Load Probability (LOLP), and Expected Unserved Energy (EUE). Among them, the EUE metric captures both frequency and magnitude of Loss of Load (LOL) events and yields intuitive interpretations for MRI and MRIC (discussed in Section II.D). It is therefore selected for the reliability metric $M$. With EUE, in *MWh per planning period*, resource $i$'s MRI, in *hours per planning period*, is defined as

$$MRI_i \equiv -\frac{\partial EUE}{\partial C_i}|_{base-case}, \forall \text{ resource } i. \tag{5}$$

MRI denotes a resource's *total* marginal reliability impact. To capture such total impact for certain resources with multiple capacity-driven parameters, the derivative in (5) should be viewed as a *total derivative* to distinguish from the partial derivative representation of MRI components in later Section II.C. The resource MRI is evaluated at a particular point on the multi-variate EUE function, i.e., the "*base case*" that summarizes all settings of load and resources needed for evaluating EUE and its derivatives. The base case is often set to the planning criteria such as "*1-day-in-10-years*" of LOLE.

The constant $K$ in (4) has the same unit as MRI and thus can be viewed as the *reference MRI* for all resources. Following the use of perfect capacity in conventional Effective Load Carrying Capability (ELCC) calculations [17], we adopt perfect capacity's MRI as the constant $K$ in (4), i.e.,

$$K = MRI_{perfect}, \tag{6}$$

where perfect capacity represents a resource that is always available at its full capacity.

With the EUE metric and the perfect capacity reference, resource $i$'s accredited capacity in (4), denoted as $MRIC_i$, is

$$MRIC_i = C_i \cdot MRI_i / MRI_{perfect}, \ \forall \ resource \ i. \quad (7)$$

---

[2] The adequacy metric is assumed a differentiable function in this paper for the convenience of notation in deriving the MRI concept. In reality, the adequacy function may not be everywhere differentiable.

The ratio of MRIs in (7) defines the *relative MRI* (rMRI) of resource *i* with respect to perfect capacity, i.e.,

$$rMRI_i \equiv MRI_i / MRI_{perfect}, \ \forall \ resource \ i. \quad (8)$$

Note that rMRI is unitless and $MRI_{perfect}$ serves as a common reference for all resources' MRIs. Then resource *i*'s MRIC can also be expressed as

$$MRIC_i = C_i \cdot rMRI_i, \ \forall \ resource \ i. \quad (9)$$

Namely, the resource's rMRI converts its native capacity to MRIC. In the following, rMRI for capacity demand will be defined to convert the native capacity demand to MRIC demand and derive the MRIC demand curves.

*II.B MRIC Demand Curves*

With MRI-based capacity accreditation, MRIC becomes the capacity market product, i.e., both offered capacities and capacity demand should be expressed in MRIC quantities. Below we derive the MRIC demand curves from our previously developed native capacity demand curves.

A capacity demand curve reflects the marginal benefits of capacity to system adequacy at different levels of the total acquired capacity. In our previous work [13] that has been implemented in ISO-NE's capacity market, the system capacity demand curve expressed in Qualified Capacity (QC), a native capacity, is defined as

$$D_{SYS}(C_{SYS}) \equiv VOLL \cdot \left(-\frac{dEUE}{dC_{SYS}}\right), \quad (10)$$

where Value of Lost Load (VOLL) is a constant calculated from the long-term market equilibrium condition, $C_{SYS}$ represents the system demand for native capacity, i.e.,

$$C_{SYS} = \sum_i C_i, \quad (11)$$

and $\left(-\frac{dEUE}{dC_{SYS}}\right)$ reflects the marginal reliability impact of having one extra MW of capacity in the system. To evaluate $\left(-\frac{dEUE}{dC_{SYS}}\right)$, one needs to know the composition of the extra MW. In our native capacity demand curve construction, $C_{SYS}$ is composed of the base case resource mix[3], and $dC_{SYS}$ represents proportional changes to the resources in the mix, i.e.,

$$\frac{dC_{SYS}}{C_{SYS}} = \frac{dC_i}{C_i}, \ \forall \ i \in base \ case. \quad (12)$$

As a result, $\frac{dEUE}{dC_{SYS}}$ can be viewed as a directional derivative with the "*direction*" being determined by the vector of native capacities composing the base case.

With MRIC being the market product, capacity demand needs to be measured in MRIC instead of native capacity $C_{SYS}$. Following the MRI-based resource accreditation, one can similarly define the MRI for system capacity demand $C_{SYS}$ as

$$MRI_{SYS} \equiv -\frac{dEUE}{dC_{SYS}}\Big|_{base-case}, \quad (13)$$

where the directional derivative is evaluated at the same base case in resource MRI definition (5). The system demand rMRI can be defined as

$$rMRI_{SYS} \equiv MRI_{SYS} / MRI_{perfect}, \quad (14)$$

and the MRIC demand can be expressed as

$$MRIC_{SYS} = C_{SYS} \cdot rMRI_{SYS}. \quad (15)$$

With (13)-(15), it can be shown in a similar way as in Section II.A that one MW of MRIC demand has the same reliability benefit as one MW perfect capacity. Thus, the *substitutability between MRI-based capacity supply and demand* quantities is achieved. Eq. (15) also indicates that the native capacity demand $C_{SYS}$ can be converted to the system MRIC demand by the adjustment factor $rMRI_{SYS}$, which can be calculated from resource rMRIs as follows.

With the vector of native capacities $[C_1, \cdots, C_i, \cdots]$ in the base case and its L1 norm $C_{SYS}$ in (11), the directional derivative in (13) can be expressed as

$$\frac{dEUE}{dC_{SYS}} = \sum_i \left(\frac{\partial EUE}{\partial C_i} \cdot \frac{C_i}{C_{SYS}}\right), \quad (16)$$

where the derivatives on both sides are taken at the same base case, allowing the substitution of (5) and (13) into (16) to yield

$$MRI_{SYS} = \sum_i \left(MRI_i \cdot \frac{C_i}{C_{SYS}}\right). \quad (17)$$

By dividing both sides of (17) with the same reference of $MRI_{perfect}$ and substituting $C_{SYS}$ with (11), we have

$$rMRI_{SYS} = \frac{\sum_i (rMRI_i \cdot C_i)}{\sum_i C_i}. \quad (18)$$

Namely, the rMRI of system capacity demand is the native capacity *weighted average rMRI* of the base case resource mix. This result is consistent with the assumption (12) in native capacity demand curve construction that one extra MW capacity demand is made of the base case resource mix.

The system MRIC demand curve $\widehat{D}_{SYS}(\cdot)$ represents the marginal reliability benefit of system MRIC capacity, i.e.,

$$\widehat{D}_{SYS}(MRIC_{SYS}) \equiv VOLL \cdot \left(-\frac{dEUE}{dMRIC_{SYS}}\right). \quad (19)$$

With the linear relationship (15) between $C_{SYS}$ and $MRIC_{SYS}$, the above MRIC demand curve (19) can be derived from the native capacity demand curve (10) as follows:

$$\widehat{D}_{SYS}(MRIC_{SYS}) = VOLL \cdot \left(-\frac{dEUE}{d(rMRI_{SYS} \cdot C_{SYS})}\right)$$

$$= \frac{1}{rMRI_{SYS}} \cdot VOLL \cdot \left(-\frac{dEUE}{dC_{SYS}}\right)$$

---

[3] A "resource mix" refers to a set of resources characterized by their shares of the total native capacity, i.e., a resource mix of *N* resources is represented by their normalized native capacities vector $[\frac{C_1}{\sum_i C_i}, \cdots, \frac{C_N}{\sum_i C_i}]$.

$$= \frac{1}{rMRI_{SYS}} D_{SYS}(C_{SYS})$$

$$= \frac{1}{rMRI_{SYS}} D_{SYS}\left(\frac{MRIC_{SYS}}{rMRI_{SYS}}\right). \quad (20)$$

The above (20) reveals that the system *MRIC* demand curve can be obtained by scaling the coordinates of the system native capacity demand curve (10) using the system demand rMRI. Similarly, a zonal MRIC demand curve can be transformed from the corresponding zonal native capacity demand curve using the zonal demand rMRI (i.e., the native capacity weighted average rMRI of resources modeled in the zone). In summary, the MRIC demand curves can be transformed from our previously developed native capacity demand curves using the corresponding system or zonal demand rMRI.

*II.C Resource MRI Calculations*

Depending on resource models, the calculation of resource MRI may take different forms. In this section, the formula for calculating MRIs of *thermal*, *energy-limited*, *intermittent* and *group* resources are presented.

*II.C.1 Calculating MRI for Thermal Resources*

A thermal resource is typically modeled in RAA as a Markov process [18]-[19]. The capacity associated with each Markov state is proportional to the resource's native capacity. The MRI of thermal resource *i* is then calculated by measuring the system EUE change with respect to a small change to its native capacity $C_i$, or equivalently proportional changes to the resource's capacities in all states, i.e., $\frac{dc_{i,s}}{c_{i,s}} = \frac{dc_i}{c_i}, \forall s$, where $c_{i,s}$ represents the resource's capacity level associated with state *s*. Then the MRI of thermal resource *i* is represented by

$$MRI_i \equiv -\frac{\partial EUE}{\partial c_i} = \sum_s \left(-\frac{\partial EUE}{\partial c_i} \cdot \frac{dc_{i,s}}{dc_i}\right)$$

$$= \sum_s \left(-\frac{\partial EUE}{\partial c_i} \cdot \frac{c_{i,s}}{c_i}\right), \forall \text{ thermal } i. \quad (21)$$

*II.C.2 Calculating MRI for Energy Limited Resources*

An energy limited resource (ELR) is typically modeled with limited energy for supporting its capacity. Consequently, the ELR's MRI is affected by both its capacity and energy limits. Suppose that an energy-limited resource *i* is modeled with its native capacity of $C_i$ MW (e.g., maximum dispatching power) and an energy limit of $E_i$ MWh. Then the marginal reliability impacts of the resource's capacity and energy limits, respectively, can be defined as:

$$MRI_{i,C} \equiv -\frac{\partial EUE}{\partial C_i} \text{ and } MRI_{i,E} \equiv -\frac{\partial EUE}{\partial E_i}. \quad (22)$$

The above $MRI_{i,C}$ and $MRI_{i,E}$, respectively, can be viewed as *capacity and energy components* of the ELR's MRI. To reflect the total marginal reliability impact of both capacity and energy limits, the ELR's capacity and energy limit are perturbed proportionally, i.e., $\frac{dE_i}{E_i} = \frac{dC_i}{C_i}$, in calculating its MRI, i.e.,

$$MRI_i \equiv -\frac{dEUE}{dC_i} = -\frac{\partial EUE}{\partial C_i} - \frac{\partial EUE}{\partial E_i} \cdot \frac{dE_i}{dC_i}$$

$$= -\frac{\partial EUE}{\partial C_i} - \frac{\partial EUE}{\partial E_i} \cdot \frac{E_i}{C_i}, \quad (23)$$

where "*d*" represents the total derivative of resource MRI and "*∂*" represents the partial derivative for MRI components. Note that for a resource *i* with unconstrained energy, $\frac{\partial EUE}{\partial E_i} = 0$ and $MRI_i$ will only reflect the marginal impact of capacity.

Substituting (22) into (23), we have

$$MRI_i \cdot C_i = MRI_{i,C} \cdot C_i + MRI_{i,E} \cdot E_i. \quad (24)$$

As will be discussed in later Section II.D, the term "$MRI_i \cdot C_i$" can be interpreted as resource *i*'s total contribution to adequacy. Then the above (24) indicates that a storage resource's contribution is made of its capacity and energy contributions.

*II.C.3 Calculating MRI for Intermittent Resources*

An intermittent resource such as wind and solar has varying outputs over time and is often modeled as hourly profiles. Consequently, the hourly outputs of the intermittent resource determine its reliability contribution. Supposed that an intermittent resource *i* has an hourly output profile $\{C_{i,t}\}_{t=1,\cdots,T}$, where $C_{i,t}$ is its MW output in hour *t* of the planning horizon of *T* hours. Define the marginal reliability impact of the resource's output in hour *t* as

$$MRI_{i,t} \equiv -\frac{\partial EUE}{\partial c_{i,t}}, \forall t = 1, \cdots, T. \quad (25)$$

The intermittent resource's MRI reflects the system EUE change with respect to a small change to its native capacity $C_i$, or equivalently proportional changes to the resource's output in all hours, i.e., $\frac{dc_{i,t}}{c_{i,t}} = \frac{dc_i}{c_i}, \forall t$. Then we have

$$MRI_i \equiv -\frac{\partial EUE}{\partial c_i} = \sum_{t=1}^{T}\left(-\frac{\partial EUE}{\partial c_{i,t}} \cdot \frac{c_{i,t}}{c_i}\right). \quad (26)$$

Substituting (25) into (26), we have

$$MRI_i \cdot C_i = \sum_{t=1}^{T}(MRI_{i,t} \cdot C_{i,t}) \quad (27)$$

With the MRI and MRIC interpretations to be discussed in later Section II.D, the above (27) indicates that an intermittent resource's reliability contribution is the sum of its contributions in individual hours.

*II.C.4 Calculating MRI for a Resource Group*

There are situations that may require calculating the MRI for a group of resources, e.g., the system capacity demand MRI in (13), or a class MRI for a technology type. Consider a group of resources with each member resource *i*'s native capacity being $C_i$ and its marginal reliability impact being $MRI_i$. Denote the group's native capacity as $C_g = \sum_i C_i$ to represent the native capacity of the group. Following the MRI definition (5), the group's marginal reliability impact $MRI_g$ can be defined as

$$MRI_g \equiv -\frac{\partial EUE}{\partial c_g}. \quad (28)$$

The above group's MRI reflects the marginal impact of the group size $C_g$. Varying $C_g$ is equivalent to proportional variations to member resources' native capacities, i.e., $\frac{dC_i}{C_i} = \frac{dC_g}{C_g}, \forall\, i \in g$. Then we have

$$MRI_g \equiv -\frac{\partial EUE}{\partial C_g} = \sum_{i\in g}\left(-\frac{\partial EUE}{\partial C_i} \cdot \frac{C_i}{C_g}\right). \quad (29)$$

With individual resource MRIs defined in (5), the above (29) can be represented as

$$MRI_g \cdot C_g = \sum_i (MRI_i \cdot C_i). \quad (30)$$

With the MRI and MRIC interpretations to be discussed in later Section II.D, the above (30) indicates that the group's reliability contribution is the sum of contributions from individual resources in the group.

*II.D MRI and MRIC Interpretations*

Below we discuss the interpretations of a resource's MRI and MRIC by introducing the concept of *MRI hours* first. In probabilistic hourly RAA simulations, each simulated scenario of the planning period is composed of hourly load and each resource's hourly available capacity. The capacity margin for each hour of the scenario can be calculated by comparing the load and total available capacity. The Unserved Energy (UE) under a scenario is the total shortage across all the hours with negative capacity margins or Loss of Load (LOL), and EUE is the probability-weighted average UE across all scenarios.

A resource MRI in (5) measures how a small increase to the resource's size, denoted by its native capacity, affects its available capacity in each hour of a simulated scenario and thus the system EUE. Depending on the hour's capacity margin, small changes to the resource's available capacity in that hour may or may not affect EUE. For an hour with a positive margin, small change to available capacity will not affect EUE. For an hour with negative margin, additional available capacity will always reduce the LOL of the hour by the same amount. For an hour with zero margin, additional available capacity *may* reduce the UE in the presence of ELRs, e.g., the extra available capacity for the hour can be shifted from that hour to the subsequent LOL hours to reduce the UE of the scenario. All the hours affecting the EUE and thus resource MRIs are termed "*MRI hours*," which include *all LOL hours* and *certain zero-margin hours*[4]. Note that MRI hours are scenario dependent.

With MRI hours, a small $\Delta$ MW increase to resource $i$'s size $C_i$ could result in extra available capacity (and thus UE reduction) for those MRI hours when the resource is available. Therefore, $MRI_i$ *can be interpreted as the expected number of MRI hours when resource $i$'s capacity is available.* In other words, $MRI_i$ measures the resource's capacity performance during MRI hours. As a special case, perfect capacity is always fully available, and thus $MRI_{perfect}$ *represents the expected number of MRI hours.* Consequently, *resource $i$'s rMRI represents its expected availability during MRI hours.*

Since resource $i$'s MRI represents its expected number of MRI hours with available capacity $C_i$, then $(C_i \cdot MRI_i)$ *represents resource $i$'s expected energy output during the MRI hours.* Consequently, $(C_i \cdot MRI_i)/MRI_{perfect}$, i.e., $MRIC_i$, can be interpreted as resource $i$'s *expected energy contribution during an MRI hour*. Note that the above MRIC interpretation of energy contribution during the common set of MRI hours implies MRIC substitutability since energy is substitutable.

*II.E Properties of MRI-based Accreditation*

Besides the substitutability property in Section II.A, the MRI-based framework has several other properties including additivity, homogeneity, common MRI hours, resource mix dependency, demand benefits preservation, and reference MRI independence. These properties are discussed as follows.

<u>*Additivity and homogeneity*</u>. *Additivity* means the accreditation value of a group of resources is equal to the sum of accreditation values of individual resources in the group, and *homogeneity* means the accreditation value of *n* identical resources is *n* times the accreditation value of an individual one. Based on the group MRI definition (28), dividing (30) by the reference $MRI_{perfect}$ yields the additivity property. Moreover, applying additivity to *n* identical group resources would lead to homogeneity, and thus homogeneity also holds for the MRI-based accreditation.

<u>*Common MRI hours*</u>. Based on the MRI and MRIC interpretations in Section II.D, the MRI hours affecting system adequacy are used to measure resources' accreditation values. Namely, the MRI accreditation employs a common set of MRI hours for evaluating all resource MRIs. The use of common MRI hours for all resources contrasts with the average accreditation scheme as will be discussed in later Section III.

<u>*Resource mix dependency*</u>. MRI is defined as a derivative of EUE. As system EUE is affected by resource mix, so is the MRI and thus the MRIC of any resource. Namely, a resource's MRI accreditation value depends not only on its own physical characteristics but on the resource mix as well. Such resource mix dependency allows the accreditation value to capture the diversity benefits of resource mix, e.g., adding capacities with similar characteristics tends to reduce their marginal reliability benefit and consequent accreditation values.

<u>*Demand benefits preservation*</u>. As discussed in Section II.B, capacity demand curves are administratively constructed to reflect the marginal reliability benefits of capacity under the base case resource mix assumption. The reliability benefits of a given size and mix of physical resources, represented by the area under a demand curve, should not depend on how resources are accredited. Such demand benefit preservation is maintained between the capacity demand curve in native capacity and the corresponding MRIC demand curve, since the latter is constructed by scaling the two coordinates of the former with reciprocal factors based on (20).

<u>*Reference MRI independence*</u>. The MRI-based accreditation uses perfect capacity MRI as the reference to define each resource's rMRI. From the derivation of the general accreditation formula (5), however, the reference $Const.$ can be an arbitrary constant while still satisfying substitutability. With a different reference (e.g., a non-perfect resource MRI), each resource's rMRI and thus MRIC will be scaled by a factor $\beta$ of the two references, i.e., $\beta = MRI_{other-ref}/MRI_{perfect}$. Consequently, each resource's offer curve will have its MRIC quantity axis scaled by $1/\beta$ and the offer price scaled by $\beta$, preserving each resource's total capacity cost under the two different references. And the demand curves will have its total MRIC quantity axis scaled by $1/\beta$ and the marginal benefit scaled by $\beta$. With the demand curves and each supply curve scaled by the same factor, the auction clearing will yield the same set of optimal physical resources and the same capacity payment for each resource, although the cleared resource quantities will be scaled by $1/\beta$ and the clearing prices will be scaled by $\beta$. In sum, the optimal resource mix and resource payments are independent of the choice of reference MRI. This reference-independent property of MRI-based accreditation indicates the importance of relative accreditation values instead of absolute ones for resources.

---

[4] Identifying zero-margin MRI hours is not a trivial task. One may numerically examine each zero-margin hour of every simulated scenario by assessing the EUE impact of adding a small capacity to that hour, which could be computationally demanding due to the large number of simulation scenarios.

*II.F Market Efficiency Under MRI-based Accreditation*

This section provides insight into how MRI-based accreditation, as compared to native capacity, could better align a capacity market with system adequacy and thus allow more efficient market outcomes.

Prior to the recent capacity accreditation reforms, most of the RTOs equipped with a capacity market procure an *adequate* amount of native capacity, i.e., capacity requirement, to meet the region's adequacy target. This is represented by the supply-demand constraint in capacity market, e.g.,

$$\sum_i C_i \geq C_{SYS}, \quad (31)$$

where $C_i$ is the native capacity of resource $i$, and $C_{SYS}$ is the capacity requirement which represents the size of a given resource mix that, together with a given load, would meet the adequacy criteria such as 0.1 days/year LOLE. The calculation of $C_{SYS}$ requires load and resource mix assumptions, which are captured in the *base case*. Note that capacity demand curve is an extension of the capacity requirement by allowing different $C_{SYS}$ levels that correspond to different system adequacy levels.

Constraint (31) does not distinguish different resources' native capacities in meeting the capacity demand. Rather, it treats system adequacy as a single-variable linear function of the total capacity $\sum_i C_i$. However, system adequacy is a multivariate nonlinear function of individual resources (represented by the resource mix vector and its size). Thus, the linear capacity requirement constraint (31) can be viewed as an approximation to the actual adequacy need. This is illustrated in Fig. 1 with a simple mix of two resources. In the figure, the actual adequacy level curve represents all the pairs $(C_1, C_2)$ that yield the same adequacy level as the at-criteria base case. The curve is in general nonlinear[5] and tends to be flatter (or steeper) as $C_1$ (or $C_2$) increases since it would require more capacity from an abundant resource to replace the other resource to maintain the same system adequacy level. The native capacity market implied level curve represents the pairs $(C_1, C_2)$ that yield the same total capacity requirement $C_{SYS}$ in the market constraint (31). The curve is linear with the slope of -1, indicating the capacity market's approximation of system adequacy need with the total capacity requirement. The two curves represent the same adequacy level as the base case, which corresponds to their intersection point. The difference between the two curves reflects the approximation error of capacity requirement to the actual adequacy need.

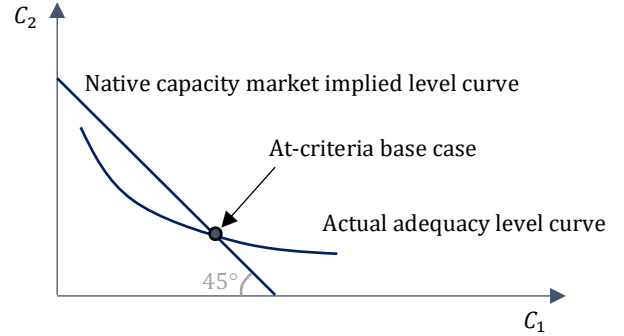

Fig. 1. Level curves of adequacy function and its approximation

With the MRI-based capacity product, both capacity supply and demand are denominated in MRIC. Thus, the native capacity requirement constraint (31) is replaced by

$$\sum_i MRIC_i \geq MRIC_{SYS}. \quad (32)$$

where $MRIC_i$ and $MRIC_{SYS}$, respectively, are individual resource MRICs and the system MRIC requirement. To compare the above MRIC requirement (32) with the native capacity requirement (31), the MRICs are substituted by (9) and (15) to yield

$$\sum_i (rMRI_i \cdot C_i) \geq rMRI_{SYS} \cdot C_{SYS}. \quad (33)$$

Constraint (33) attaches rMRIs as weights to individual resources in meeting the capacity requirement. Namely, the same amount of native capacity from different resources are not treated as substitutable. The implied adequacy level curve for the above two-resource example is depicted in Fig.2. In the figure, the actual adequacy level curve remains unchanged from Fig.1 as it reflects the physical characteristics of the system. The MRIC market implied curve represents the pairs $(C_1, C_2)$ that yield the same total MRIC requirement $MRIC_{SYS}$ in the market constraint (32). The curve is linear and intersects with the actual adequacy level curve at the point corresponding to the base case when resource rMRIs and demand $rMRI_{SYS}$ are calculated from the same base case. The slope of the curve is determined by the vector $[rMRI_1, rMRI_2]$, which represents the gradient of the actual adequacy level curve at the base case point when EUE is used as the adequacy measure[6]. Thus, the adequacy level curve implied by the MRIC requirement constraint (33) can be viewed as a linear approximation to the nonlinear adequacy level curve. By incorporating the gradient information of adequacy function at the base case, the MRI-induced linear approximation is tangent to the actual nonlinear adequacy curve and thus is more accurate than the one under native capacity *near the base case point*. Assume that the *optimal solution* under the actual nonlinear adequacy function is not far from the base case point, which likely holds in practice since the base case is often constructed from the set of existing or recently cleared resources and a drastic change from the set is typically costly and thus unlikely to be optimal. Then, a more accurate approximation near the base case would likely lead to a market solution closer to the optimal one.

---

[5] The actual adequacy level curve can be linear when both resources are perfectly 1:1 substitutable, e.g., both are perfect capacity.

[6] Note that the use of EUE to characterize system adequacy will not contradict the adequacy criteria defined on a different adequacy measure, e.g.,

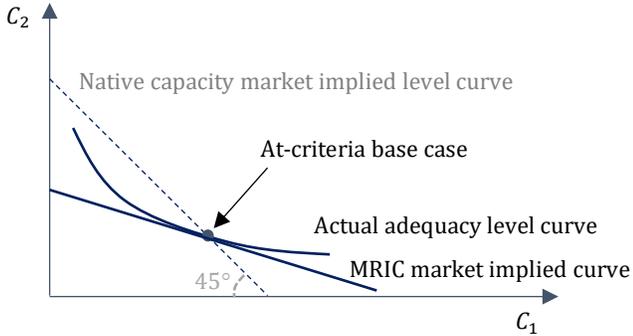

Fig. 2. Level curves of adequacy function and its MRIC approximation

The above analysis shows that the MRIC requirement (32) provides a more accurate approximation to the nonlinear adequacy function than the native capacity requirement (31). This is reflected in more accurate representations of individual resources' reliability contributions (i.e., accredited capacities) and a more accurate model of the reliability function (i.e., EUE), thus allowing more efficient capacity market outcomes.

### III. COMPARISON WITH OTHER ACCREDITATION METHODS

In this section, existing accreditation methods of ICAP/UCAP (Section III.A), Average ELCC (Section III.B) and Marginal ELCC (Section III.C) are introduced and compared with the MRI based accreditation.

*III.A ICAP/UCAP*

Prior to the recent wave of capacity accreditation reform, ISO/RTOs often define a resource's accreditation based on its ICAP or UCAP. For example, ISO-NE and IESO use the ICAP concept for their capacity accreditations while PJM and NYISO use the UCAP concept. The main difference between ICAP and UCAP is: The ICAP of a resource does not reflect its outage information, while the UCAP of the resource applies an outage-rate-based discount to the ICAP.

Both ICAP and UCAP of a resource are *determined only by the resource's own characteristics*. As a result, ICAP or UCAP does not provide an accurate characterization of a resource's reliability contribution and is generally *not substitutable* between different resources, since system reliability metrics such as LOLE and EUE are inseparable functions in individual resources and one resource's reliability contribution is affected by other resources.

In contrast, the MRIC of a resource is determined by not only the resource's own characteristics, but load and other resources as well. Such resource mix dependence allows diversity benefits to be reflected in resource accreditation values.

*III.B Average ELCC*

The *Average* Effective Load Carrying Capability[7] (AELCC) method (conventionally known as ELCC method) accredits a resource by the amount of perfect capacity that can replace the resource for the same level of system reliability. LOLE is typically adopted as the reliability metric in ELCC calculations. The AELCC method can also be applied to a class of resources by replacing the entire class with an equivalent amount of perfect capacity.

The AELCC method is illustrated in Fig. 3 for a resource (or class) $i$ with a native capacity of $C_i$ MW. The difference in system adequacy levels with and without the resource is represented by $\Delta$LOLE. In the figure, a perfect capacity of $C_{i,perfect}$ MW is assumed to replace the resource of $C_i$ MW to yield the same system LOLE impact. Note that the LOLE curves are shown as decreasing functions of capacity as additional capacity would improve system adequacy (i.e., reducing LOLE). Also, the steepness of the curves reduces with the increase of capacity, reflecting the generally decreasing marginal reliability benefit of additional capacity. The AELCC method accredits resource $i$ at $C_{i,perfect}$ MW, i.e.,

$$AELCC_i \equiv C_{i,perfect}. \tag{34}$$

For the convenience of comparing different accreditation methods, we define a resource's Contribution Factor (CF) as the ratio of its accredited capacity to its native capacity. Then for AELCC, resource $i$'s CF can be represented as

$$CF_i^{AELCC} \equiv \frac{C_{i,perfect}}{C_i} = \frac{\Delta C_{i,perfect}/\Delta LOLE}{\Delta C_i/\Delta LOLE} \tag{35}$$

where $\Delta C_{i,perfect} = C_{i,perfect} - 0$ and $\Delta C_i = C_i - 0$. The above formula indicates that to yield the same reliability impact of $\Delta LOLE$ would require $\Delta C_i$ MW of resource $i$ or $\Delta C_{i,perfect}$ MW of perfect capacity. Two RAA cases are involved in the AELCC evaluation for resource $i$: One with resource $i$ removed from the base case (represented by point $A$ in Fig. 3); and the other case with the replacement perfect capacity (point $B$). The former case evaluates the LOLE impact of resource $i$ and the latter case determines the replacement perfect capacity value (i.e., $\Delta C_{i,perfect}$) that yields the same LOLE impact[8].

---

the 1-in-10 LOLE criteria can be used to construct the base case, while EUE can be used to derive the level curves and their gradients.

[7] The word "average" is used to distinguish the method from the marginal ELCC method introduced in Section III.C.

[8] The calculation of the replacement perfect capacity value often involves multiple RAA runs and can be computationally demanding, in contrast to a single RAA run required for MRI calculation.

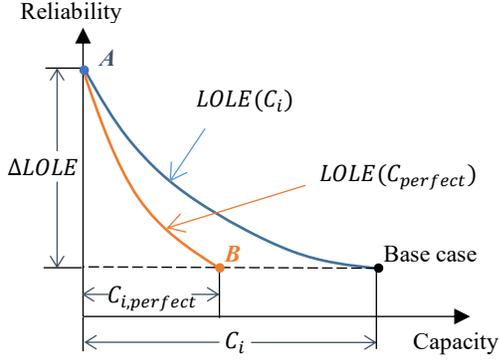

Fig.3. Illustration of AELCC method.

For each resource, the AELCC method finds a perfect capacity value that yields the same LOLE impact. As LOLE is affected by resource mix, the AELCC value of a resource is *dependent on the resource mix*. Moreover, the RAA cases involved in resource $i$'s AELCC calculation and those involved in other resources' calculations have different resource mixes, i.e., the RAA cases $A$ and $B$ in Fig.3 are resource dependent, leading to different LOL events patterns for different resources' AELCC calculations. Therefore, unlike the MRI-based accreditation characterized by common MRI hours, the AELCC is *not evaluated on a common set of hours* for different resources, implying that accredited AELCC capacities of different resources are *not substitutable in terms of reliability contribution*. Also, with nonlinear LOLE functions, *additivity and homogeneity do not hold* with the AELCC accreditation, i.e., the AELCC value of a resource group is not the sum of AELCC values of individual resources in the group, and the AELCC value of $n$ identical *resources* isn't equal to $n$ times the value of an individual resource.

*III.C Marginal ELCC*

*Marginal* ELCC (MELCC) is the marginal version of the ELCC method. The MELCC calculation can follow a similar process as the AELCC calculation, except that a small (marginal) perturbation to a resource is performed instead of the removal of the entire resource. Namely, it calculates the equivalent amount of perfect capacity that would yield the same LOLE impact of a small change to a resource's capacity.

The MELCC calculation is illustrated in Fig. 4, where the adequacy impact $\partial LOLE$ of a small change $\partial C_i$ to resource $i$, and the perfect capacity amount $\partial C_{i,perfect}$ that would yield the same adequacy impact are illustrated. Note that with the marginal changes, the LOLE functions are plotted as linear ones in the figure. Following (35), the contribution factor for the small change $\partial C_i$ of resource $i$ can be represented as

$$CF_i^{MELCC} \equiv \frac{\partial C_{i,perfect}}{\partial C_i} = \frac{\partial C_{i,perfect}/\partial LOLE}{\partial C_i/\partial LOLE} \quad (36)$$

where "$\partial$" indicates marginal changes to be distinguished from the "$\Delta$" changes of the entire resource in AELCC. The above formula indicates that to yield the same reliability impact of $\partial LOLE$ would require $\partial C_i$ MW resource $i$ or $\partial C_{i,perfect}$ MW perfect capacity. Then the MELCC of resource $i$ is defined as:

$$MELCC_i \equiv C_i \cdot CF_i^{MELCC} = C_i \cdot \frac{\partial C_{i,perfect}/\partial LOLE}{\partial C_i/\partial LOLE} \quad (37)$$

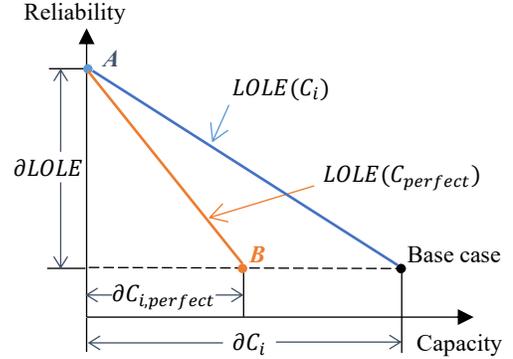

Fig.4. Illustration of MELCC method.

The CF in (36) can also be written as:

$$CF_i^{MELCC} = \frac{\partial LOLE/\partial C_i}{\partial LOLE/\partial C_{perfect}}, \quad (38)$$

which resembles the rMRI definition (8) except for the use of LOLE metric. Indeed, if EUE were adopted, MELCC would be equal to MRIC in theory. As a result, the MELCC accreditation shares the *resource-mix dependency* property with the MRI-based accreditation. The *additivity and homogeneity properties also hold* due to the marginal nature of the MELCC concept. With LOLE metric, however, a resource's *MELCC does not have the interpretation of energy contribution* as MRIC. Also, LOLE is less sensitive to capacity changes as compared to EUE or could even be discontinuous, and thus the practical MELCC calculation of LOLE derivatives may require sizable perturbations to incur traceable LOLE changes, undermining the marginal nature of the method.

IV. NUMERICAL TESTING

In this section, we test different accreditation methods on a 25-unit system with GE-MARS [20] and examine their properties discussed in the previous sections. A planning period of 8760 hours (i.e., one year) is considered for a system with 10 probabilistic hourly load profiles built on a historical year's load shape in New England, 12 thermal units of different sizes and outage rates (each unit modeled as a two-state Markov Chain with the "1" state representing full capacity and "0" representing zero capacity), 11 Intermittent Power Resources (IPRs) (each modeled as probabilistic hourly output profiles), 2 Energy Storage (ES) units (each modeled with charging capacity, discharging capacity and energy limit, and without forced outage or efficiency loss), and a perfect resource representing the remaining resources in the system. The ICAPs, forced outage rates (for non- ES units) and energy limits (for ES units) are listed in Table 1. For simplicity, no transmission limit or unit maintenance is considered. The system is verified to be at 0.1 days/year LOLE and thus serves as the *base case* for accreditation calculations.

Following the MRIC calculation described in Section II.B, and the average and marginal ELCC calculations described in Section III, the resulting accredited capacities under these accreditation methods are summarized in Table 1. For MRIC calculation, the perturbation size of 1 MW is applied. For AELCC calculation, a trial-and-error method is used to identify the perfect capacity value that yields the closest LOLE impact as a resource. Note that for each resource, the AELCC calculation requires multiple evaluations of candidate values and thus is computationally more expensive than the MRIC calculation that evaluates only once for the perturbed capacity. For MELCC calculation, to reduce the computational burden we apply a similar perturbation process to the MRIC calculation by examining the LOLE impact of "small" capacity changes based on (38). Noted that the LOLE function is discontinuous at some capacity values, and thus the MELCC could be undefined in theory and have numerical issues in practice.

From Table 1, the AELCC, MELCC and MRIC accreditation values of a unit are no higher than its ICAP since all these accreditation methods represent the unit's imperfect installed capacity to some "equivalent" perfect capacity. Note that for perfect capacity, all three methods yield the ICAP value as expected. Also, MELCC and MRIC, both being marginal methods, yield different accreditation values due to their adoption of different adequacy metrics of LOLE and EUE, respectively. AELCC and MELCC, both using the LOLE metric, generally yield different accreditation values due to the *nonlinearity* of the LOLE function.

Observe that MRIC of a thermal unit $i$ is no more than its UCAP value of $ICAP_i \cdot (1 - FOR_i)$, e.g., TH1 unit's MRIC of 605.6 MW is less than its UCAP value of 750×(1-0.1041) = 671.9 MW. The reason is: UCAP reflects the unit's expected availability across all hours, while MRIC reflects its expected availability during the MRI hours. As an hour is more likely to be an MRI hour when the unit is on outage than when the unit is available (with all else being the same), the unit is more likely to be unavailable during the MRI hours.

Table 1: Average ELCC, Marginal ELCC, and MRIC.

| Unit | Native $ICAP_i$ | $FOR_i$ or $E_i$ | AELCC | MELCC | MRIC |
|---|---|---|---|---|---|
| TH1 | 750 | 0.1041 | 631 | 560.6 | 605.6 |
| TH2 | 150 | 0.0097 | 148 | 148.5 | 148.4 |
| TH3 | 100 | 0.2061 | 79 | 76.8 | 77.9 |
| TH4 | 1000 | 0.008 | 986 | 969.7 | 979.6 |
| TH5 | 5 | 0.0723 | 4.6 | 4.6 | 4.6 |
| TH6 | 10 | 0.0723 | 9.8 | 9.2 | 9.3 |
| TH7 | 150 | 0.0964 | 134 | 133.3 | 133.8 |
| TH8 | 600 | 0.0381 | 567 | 551.5 | 561.0 |
| TH9 | 900 | 0.016 | 879 | 836.4 | 866.5 |
| TH10 | 50 | 0.001 | 49.5 | 50 | 49.9 |
| TH11 | 200 | 0.001 | 199 | 200 | 199.7 |
| TH12 | 300 | 0.006 | 298 | 293.9 | 297.7 |
| IPR1 | 900 | 0 | 145 | 132.9 | 138.9 |
| IPR2 | 600 | 0 | 145 | 117.3 | 133.7 |
| IPR3 | 1400 | 0 | 153 | 92.8 | 228.4 |
| IPR4 | 60 | 0 | 35 | 34.8 | 35.3 |
| IPR5 | 275 | 0 | 185 | 191.7 | 181.8 |
| IPR6 | 15 | 0 | 7.8 | 8.0 | 8.0 |
| IPR7 | 700 | 0 | 159 | 136.6 | 159.7 |
| IPR8 | 25 | 0 | 16.5 | 16.8 | 16.6 |
| IPR9 | 200 | 0 | 197 | 198 | 193.2 |
| IPR10 | 50 | 0 | 45.5 | 45.3 | 44.2 |
| IPR11 | 800 | 0.7962 | 12.9 | 15.1 | 23.6 |
| ES1 | 600 | 1200 | 570 | 515.2 | 224.0 |
| ES2 | 1500 | 12000 | 1500 | 1500 | 1437.2 |
| Perfect | 21576.3 | 0 | 21576.3 | 21576.3 | 21576.3 |
| Total | 32916.3 | - | 28911.9 | 28415.2 | 28334.9 |

The contribution factors (i.e., accreditation value divided by ICAP) of the storage units are listed in Table 2. It can be seen the contribution factor of the 2-hour storage is lower than that of the 8-hour storage for each accreditation method. This can be explained as the longer-duration storage is less likely to be energy constrained and thus acts more like a perfect capacity. The low rMRI (i.e., 0.37) of the 2-hour storage can be explained by the interpretation of rMRI as the expected availability during the MRI hours (Section II.D) and the relatively long expected duration of loss of load events (i.e., 4.2 hours) in the base case.

Table 2: Contribution factors of 2-hour and 8-hour storage units.

| Storage Unit | CF-AELCC | CF-MELCC | CF-MRIC (rMRI) |
|---|---|---|---|
| ES1 (2-hour) | 0.95 | 0.86 | 0.37 |
| ES2 (8-hour) | 1.0 | 1.0 | 0.96 |

To test the additivity of different accreditation methods, consider IPRs 1-8 as a single group. The group's accredited values under different accreditations are calculated and listed in Table 3, along with the sum of individual accreditation values in Table 1. It can be seen from Table 2 that the group AELCC (i.e., 903 MW) is 56.7 MW more than the sum of individual resource AELCCs (i.e., 846.3 MW), consistent with the non-additive feature of AELCC method. The MRIC of the group, whether calculated as a single group or the sum of individual unit MRICs, yields almost identical results with the small 0.3 MW difference attributing to numerical tolerance, which is consistent with the additive feature of MRIC. The MELCC accreditation, although additive in theory with assumed differentiability of LOLE, shows sizable gap (25.7 MW) between the group MELCC (756.6 MW) and the sum of individual MELCCs (730.9), likely caused by the numerical challenges of MELCC calculation (i.e., discontinuity of LOLE function and low LOLE sensitivity to small capacity changes).

Table 3: Group accreditation and total individual accreditations.

| IPR Group | ICAP | AELCC | MELCC | MRIC |
|---|---|---|---|---|
| Group Accreditation | 3975 | 903 | 756.6 | 902.8 |
| Sum of Individual Accred. | 3975 | 846.3 | 730.9 | 902.5 |

To examine the reference dependence, consider TH1 as the new reference. Then resources' accreditation values under the new reference can be calculated and the corresponding contribution factors can be compared with the values under the perfect capacity reference. Table 4 lists the contribution factors of two units, TH1 and TH2, under the two different references for AELCC, MELCC and MRIC methods. Note that a resource's ICAP is not affected by the reference choice. Also, when TH1 is the reference, its contribution factors are 1. The CFs of TH2 under the new reference TH1 capacity are higher than those under perfect capacity since the new reference is less perfect. Note that the CFs of TH2 exceed 1 since TH2 with a small outage rate is more likely to be available than the new reference TH1 with a much higher outage rate. The percentage increases of accreditation values under the TH1 reference with respect to those under the perfect reference are also listed in Table 4. Observe that marginal methods (i.e., MELCC and MRIC) impose a uniform percentage increase to the accreditation values of both units without affecting the *relative* capacity values between the two units, and are thus considered reference independent. AELCC, however, results in accreditation increases of 18.9% for TH1 and 12.2% for TH2, respectively, altering the relative values of the two units, i.e., TH1 gained an advantage over TH2 due to the reference shift. Therefore, AELCC is reference dependent.

Table 4: Contribution factors with different references.

| Unit | ICAP | CF-AELCC | CF-MELCC | CF-MRIC |
|------|------|----------|----------|---------|
| With perfect capacity as reference ||||| 
| TH1 | 750 | 0.841 | 0.747 | 0.807 |
| TH2 | 150 | 0.987 | 0.990 | 0.989 |
| With TH1 as reference ||||| 
| TH1 | 750 | 1.0 (+18.9%) | 1.0 (+33.8%) | 1.0 (+23.9%) |
| TH2 | 150 | 1.11 (+12.2%) | 1.32 (+33.8%) | 1.23 (+23.9%) |

## V. Conclusion

An MRI-based accreditation framework is introduced for capacity markets. The framework is analyzed for its features and compared with other accreditation methods. With the MRI-based resource accreditation and capacity demand, the capacity market of MRIC product is better aligned with the system adequacy need, allowing more efficient market outcomes.

## References

[1] NERC Reliability Terminology. [online available] https://www.nerc.com/AboutNERC/Documents/Terms%20AUG13.pdf

[2] AESO, Comparison of Installed Capacity (ICAP) & Unforced Capacity (UCAP) Capacity Value Calculation Methods, 2017, [online available] https://www.aeso.ca/assets/Uploads/20170704-Eligibility-Session-3-UCAP-ICAP-Comparison-Presentation.pdf

[3] Capacity Auction Reforms Key Projects, ISO-NE, [online available] https://www.iso-ne.com/committees/key-projects/capacity-auction-reforms-key-project

[4] Resource Accreditation, MISO white paper, May 2023, [online available] https://cdn.misoenergy.org/MISO%20Draft%20Resource%20Accreditation%20Design%20White%20Paper628865.pdf

[5] Z. T. Smith and M. Mohrman, "Improving capacity accreditation," https://www.nyiso.com/documents/20142/29177064/Capacity%20Accreditation%2003-16-22%20v7.pdf/b26e6a99-5f4e-29cc-c60c-47608c78c983, 2022, [Online; accessed 4-December-2024].

[6] C. Glazer and C. Lu, "Capacity market reforms to accommodate the energy transition while maintaining resource adequacy," PJM Interconnection LLC, Tech. Rep., 2023.

[7] ISO-NE, "Resource Capacity Accreditation in the Forward Capacity Market: MRI-based conceptual design," Jul. 2022, [Online available]

[8] ISO-NE, "Resource Capacity Accreditation in the Forward Capacity Market: continued discussion on conceptual design," Aug. 2022, [Online available]

[9] D. Stenclik, "Ensuring efficient reliability: New design principles for capacity accreditation," Energy Systems Integration Group, Tech. Rep., 2023.

[10] P. Huang, "What new reforms to capacity accreditation mean for us power markets," https://www.woodmac.com/news/opinion/what-new-reforms-to-capacity-accreditation-mean-for-us-power-markets/, 2024, [Online; accessed 4-December-2024].

[11] T. Aagaard and A. N. Kleit, "Marginal vs. average effective load carrying capability: How should capacity markets deal with alternative generation forms?", Utilities Policy, Vol. 84, Oct. 2023.
A. D. Kevin Carden and R. Amitava, "Effective load carrying capability study," Astrap´e Consulting, Tech. Rep., 2024. [online available] https://www.ercot.com/files/docs/2025/02/12/2024ERCOT_ELCC_Study_Final_Report02112025.pdf

[12] F. Zhao, T. Zheng, and E. Litvinov, "Constructing demand curves in forward capacity market," *IEEE Transactions on Power Systems*, vol. 33, no. 1, pp. 525–535, 2017.

[13] F. Zhao, T. Zheng, and E. Litvinov, "Economic interpretation of demand curves in multi-product electricity markets- part I: Theory," *IEEE Transactions on Power Systems,* vol. 35, no. 6, pp. 4880–4887, 2020.

[14] F. Zhao, T. Zheng, and E. Litvinov, "Economic interpretation of demand curves in multi-product electricity markets- part II: Practice," *IEEE Transactions on Power Systems*, vol. 35, no. 6, pp. 4888–4895, 2020.

[15] NERC, Probabilistic Assessment - Technical Guideline Document [online available] https://www.nerc.com/comm/RSTC/PAWG/proba_technical_guideline_document_08082014.pdf

[16] NERC, Methods to Model and Calculate Capacity Contributions of Variable Generation for Resource Adequacy Planning [online available] https://www.nerc.com/pa/RAPA/ra/Reliability%20Assessments%20DL/IVGTF1-2.pdf

[17] R. Billinton and R. N. Allan, Reliability Evaluation of Power Systems, Springer, 1996.

[18] C. Singh, P. Jirutitijaroen, J. Mitra, Electric Power Grid Reliability Evaluation: Models and Methods, John Wiley & Sons, 2018

[19] GE, "Multi-area reliability simulation (MARS) software," https://www.gevernova.com/content/dam/Energy_Consulting/global/en_US/pdfs/GEA35097B-GE-MARS_Brochure_R3.pdf, 2018, [Online; accessed 4-December-2024].